\documentclass[printer]{aa}
\usepackage{color}
\usepackage{newtxtext,newtxmath}
\usepackage{graphicx}
\usepackage{amssymb}
\usepackage{amsmath}
\usepackage{hyperref}
\usepackage{natbib}

\graphicspath{{./pics/}}

\def\gr{$\gamma$-ray}
\begin{document}

\title{Search for decaying eV-mass axion-like particles using gamma-ray signal from blazars}

\author{A. Korochkin \inst{1,2} \and 
A. Neronov \inst{1,3} \and
D. Semikoz \inst{1} }
 \date{}

\institute{APC, Universite Paris Diderot, CNRS/IN2P3, CEA/IRFU \email{alexander.korochkin@apc.in2p3.fr} \and
Institute for Nuclear Research of the Russian Academy of Sciences,
		60th October Anniversary st. 7a, 117312, Moscow, Russia \and
Astronomy Department, University of Geneva, Ch. d'Ecogia 16, 1290, Versoix, Switzerland}

\label{firstpage}

\abstract
{
Decaying axion-like particles (ALP) with masses in the eV range which might occupy dark matter halos of the Milky Way and other galaxies  produce a characteristic "bump" feature in the spectrum of extragalactic background light (EBL). This feature leaves an imprint on the gamma-ray spectra of distant extragalactic sources.  }
{
We derive constraints on the ALP coupling to photons  based on analysis of spectra  of very-high-energy gamma-ray loud blazars.
} 
{
We combine gamma-ray spectral measurements by Fermi/LAT and Cherenkov telescopes and fit a model in which the intrinsic source spectrum is modified by pair production on photons produced by ALP decays. We constrain the amplitude of gamma-ray flux suppression by this effect. 
}{ 
We find that the combined Fermi/LAT and VERITAS data set for the source  1ES 1218+304 currently provides the tightest constraint on ALP-two-photon coupling which is complementary to the constraints imposed by non-observation of excess energy loss in Horizontal Branch stars, by the high-resolution spectroscopic observations of galaxy clusters with optical telescopes and by the searches of ALP signal with CERN Solar Axion Telescope. Our analysis favours existence of a bump in the EBL spectrum which could be produced by ALPs in the mass range 2-3 eV and axion-photon coupling $\sim 10^{-10}$ GeV$^{-1}$. We discuss possibilities for verification of this hint  with deeper Cherenkov telescope observations of large number of blazars  with current generation instruments and with the Cherenkov Telescope Array. 
}
{}


\maketitle

\section{Introduction}

\begin{figure}
	\centerline{
		\includegraphics[width=1\linewidth]{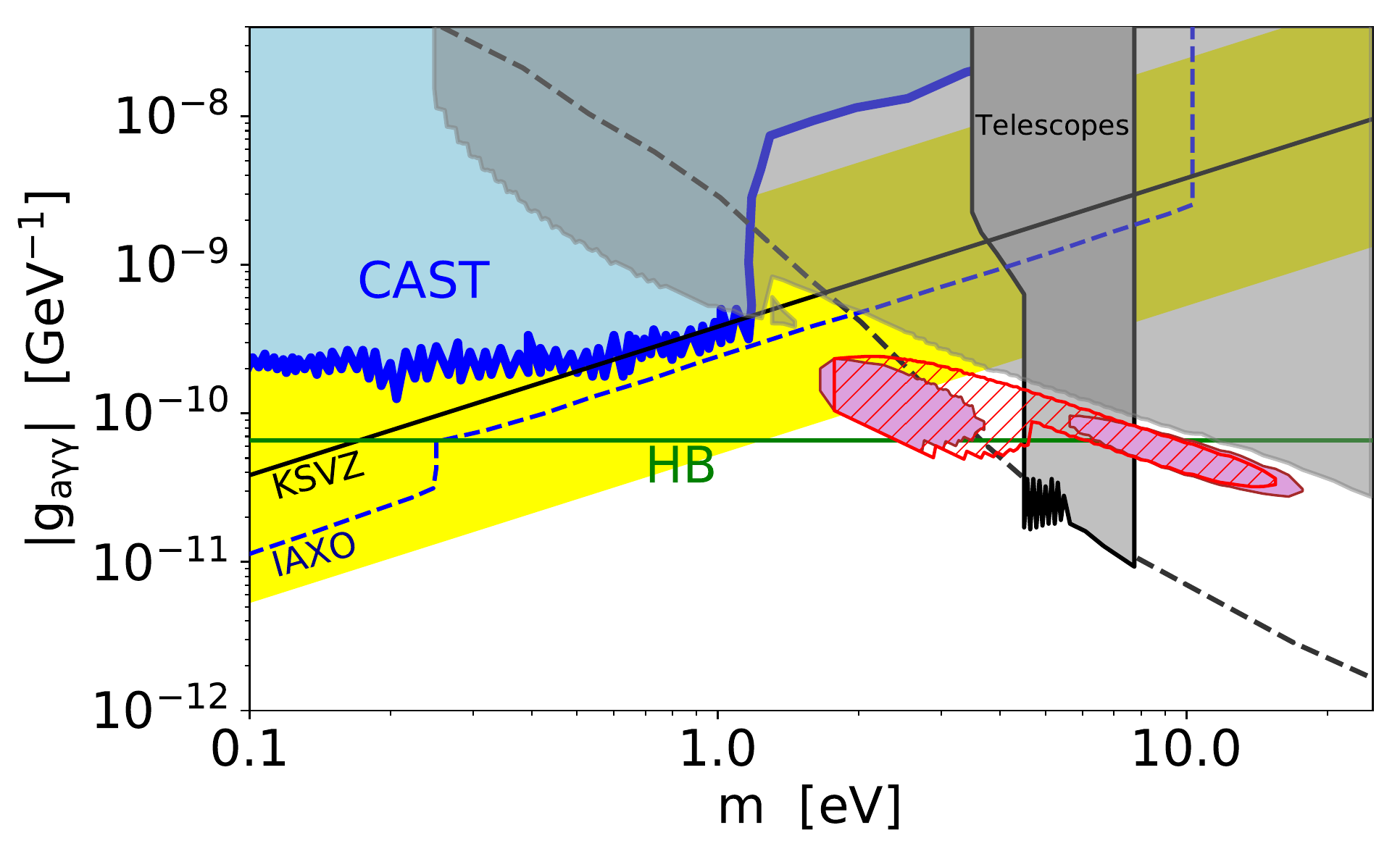}
	}
	\caption{\label{fig:eblalp} Constraints of ALP parameters from laboratory and astrophysical probes.  			Yellow band and black solid line corresponds to the  QCD 
			 axion models \citep{PhysRevLett.43.103,SHIFMAN1980493,DINE1981199,Zhitnitsky:1980tq}. Blue region is the limit impost by CAST non-observation of solar axions \citep{Arik:2013nya}. Grey vertical
			 band comes from high-resolution spectroscopy of intracluster medium in by optical telescope  \citep{Grin:2006aw}. Black dashed line is the EBL limit estimate of \citet{Arias:2012az}. Green horizontal line is the limit imposed by non-observation of excess energy loss in Horizontal Branch stars \citep{Straniero:2015nvc}. Light grey shaded region shows 95\% confidence level excluded range of parameters from non-observation of absorption feature in the Fermi/LAT+VERITAS spectrum of 1ES 1218+304 (discussed in this paper). Violet shaded ellipses show 68\% confidence level preferred range of the parameters obtained by the spectral analysis of 1ES 1218+304. Red hatched region shows the favoured region found if the first spectral point of VERITAS is not taken into account.}
\end{figure}

Axion-like particles (ALPs) are hypothetical neutral light bosons with zero spin which arise in many extensions of the Standard Model of particle physics (see e.g. \citet{2010ARNPS..60..405J,review} for reviews). They are the generalisation of the QCD axions which initially were invoked to solve the strong CP problem \citep{PhysRevLett.38.1440,PhysRevLett.40.223,PhysRevLett.40.279}. 
ALPs are characterised by the mass $m$ and two-photon coupling constant $g_{a\gamma\gamma}$, the parameter space shown in Fig. \ref{fig:eblalp}. These two parameters are expressed through one and the same (unknown) energy scale for the conventional axion, so that they are related as\footnote{We use Natural system of units in which the Planck constant and the speed of light are $\hbar=c=1$}. $g_{a\gamma\gamma}\simeq 2\times 10^{-10}\zeta(m_a/1\mbox{ eV}))$~GeV$^{-1}$  ($\zeta$  is a model-dependent parameter of the order of  one) \citep{PhysRevLett.43.103,SHIFMAN1980493,DINE1981199,Zhitnitsky:1980tq}. An example of this relation for a particular Kim-Shifman-Vainshtein-Zakharov (KSVZ) model is shown by black line and yellow model uncertainty range in Fig. \ref{fig:eblalp}. This constraint is relaxed in generic ALP models, so that broad region of $(m,g_{a\gamma\gamma})$ parameter space could be considered \citep{2010ARNPS..60..405J}.

 The ALPs provide a viable dark matter candidate \citep{Arias:2012az}. They could have been produced in the Early Universe through misalignment of initial field values of the ALP scalar field at the onset of spontaneous breaking of symmetry associated to the ALP field.  Their coupling to photons leads to gradual disappearance of the relic ALPs through two-photon decays on the time scale $\tau\simeq 10^{25} \left(g_{a\gamma\gamma}/10^{-10}\mbox{ GeV}^{-1}\right)^{-2}\left(m/1\mbox{ eV}\right)^{-3}$~s, so that ALPs with mass $m>1\left(g/10^{-6}\mbox{ GeV}^{-1}\right)^{-2/3}$~eV decay on the time scale shorter than the age of the Universe. Smaller mass axions survive till present epoch and could form the dark matter. In spite of their very low mass, in the present-day Universe ALPs form cold, rather than hot, dark matter because of collective effects of Bose-Einstein condensation.

The two-photon vertex interactions determine the preferred method of search for ALPs by their conversion to photons in the presence of magnetic fields \citep{PhysRevLett.51.1415}. 
Non-observation of this effect imposes constraints on the ALP parameters in the $m-g_{a\gamma\gamma}$ parameter space, based on laboratory experiments and astronomical observations. Non-observation of axion flux from the Sun by CAST experiment constrains $g_{a\gamma\gamma}$ from above in the mass range below 1~eV (see Fig. \ref{fig:eblalp}) \citep{Arik:2013nya}. The CAST limit will be improved by the next-generation facility  IAXO \citep{iaxo}, also shown in Fig. \ref{fig:eblalp}.   

A strong constraint on the two-photon coupling in a broad mass range is imposed by non-observation of the effect of energy loss through emission of ALPs on stellar evolution \citep{PhysRevD.33.897,PhysRevLett.113.191302,Straniero:2015nvc}. This constraint, shown by the horizontal line in Fig. \ref{fig:eblalp}, stems from non-detection of excessive energy loss through ALP emission which would affect evolution of the Horizontal Branch (HB) stars in globular clusters. Similarly to most of the astrophysical constraints, the stellar evolution argument relies on the assumption of validity of complex stellar evolution models controlled by large number of parameters which are not directly measured \citep{second_parameter_problem,third_parameter}. 

Direct searches for ALP decays into two photons  were performed using high-resolution spectroscopy of galaxy clusters  in the range of masses between 4.5 and 7.7 eV  \citep{Grin:2006aw}. Non-observation of non-identified spectral emission lines from ALP decays imposes a constraint on $g_{a\gamma\gamma}$ shown by grey shading in Fig. \ref{fig:eblalp}.  

Line emission from ALPs decaying everywhere across the Universe also contributes to the Extragalactic Background Light (EBL). Superposition of the line emission accumulated at different redshifts leads to a broad "bump-like" feature in the EBL spectrum with the width $\Delta E\sim E$. Such feature could be directly measured by dedicated observations probing the EBL. Such measurements \citep{Wright:1999zb,Matsumoto:2004dx,Tsumura:2013iza,Matsumoto:2015fma,Matsuura:2017lub} are, however, challenging in the eV photon energy range because of the presence of strong Zodiacal light background. The direct measurements typically find an excess EBL flux above the estimate from direct galaxy counts in the near-infrared \citep{Xu:2004zg,Madau:1999yh,Keenan:2010na,Fazio:2004kx}. Recent measurements by CIBER \citep{Matsuura:2017lub} estimate the excess flux  at the level 29 to 42 $nW/m^2/sr$, depending on assumptions about Zodiacal flux model. Recent results from AKARI \citep{Tsumura:2013iza} between $\sim 2 \mu m$ and $\sim 5 \mu m$ also demonstrate an increase in EBL intensity at shorter wavelengths. 

Attenuation of gamma-ray signal  from distant sources by the effect of pair production on the EBL photons  provides an alternative way of probing the EBL  \citep{Ahnen:2016gog,Abramowski:2012ry,desai,Ajello:2018sxm,Acciari:2019zgl}. This method has been used to set limits on the normalisation of the EBL, under certain assumptions about its spectral shape.  However, such an analysis does not account for possible unforeseen spectral features in the EBL, such as e.g. the ALP dark matter decay bump. 

It is nevertheless possible to include additional EBL model components in the $\gamma$-ray data analysis. In a recent study \citep{eblbump}  we have combined Fermi/LAT and Cherenkov telescopes  $\gamma$-ray spectra of a large number of blazars and derived tight constraints on possible EBL bump features by fitting the $\gamma$-ray spectra with models which include additional attenuation of the $\gamma$-ray flux by absorption on the EBL bump. Presence of the bump leads to a characteristic dip in the observed $\gamma$-ray spectra of distant sources. We have shown that account of possible spectral feature in the EBL makes  constraints on the EBL flux from the $\gamma$-ray measurements  consistent with "minimal EBL" flux level reported by direct measurement experiments in the 1 micron wavelength range \citep{Matsuura:2017lub}.  

In what follows we discuss implications of this result for the ALP dark matter model. We develop the line of reasoning of \citet{Arias:2012az} on constraints on the ALP two-photon coupling imposed by non-observation of the features in the EBL spectrum with a qualitative analysis (the limits on $g_{a\gamma\gamma}$ derived by \citet{Arias:2012az} have order-of-magnitude nature). The analysis of \citet{eblbump} has favoured non-zero normalisation of the EBL excess at $\sim 1$ micron wavelength. This could be directly expressed in terms of the ALP model of the EBL bump feature. The favoured bump wavelength and normalization parameters could be converted into the favoured ALP coupling and mass range. We discuss a possibility of scrutinising this result with deeper observations of a set of sources with current and next generation $\gamma$-ray telescopes.

\section{EBL spectral feature produced by decaying ALPs}

Dark matter decays at redshift $z$ produce signal at wavelengths $\lambda=\lambda_0(1+z)$, where longer than $\lambda_0=4\pi / m$ is the wavelength of ALP decay  line at the redshift $z=0$. We assume monochromatic decay line for dark matter with wavelength $\lambda_0$:
\begin{equation}
	F_{\lambda}(\lambda) = 2\frac{h c}{\lambda_0^2} \Gamma \delta\Big(\frac{\lambda}{\lambda_0} - 1\Big)
\end{equation}
where $\Gamma$ is the decay rate into two photons
\begin{equation}
	\Gamma = \frac{g_{a\gamma\gamma}^2 m^3}{64\pi}
\end{equation}

The flux generated by dark matter decays is \citep{2006MNRAS.370..213B}
\begin{equation}
\label{eq:i}
	I_{\lambda}(\lambda) = \frac{c}{4\pi} \int\limits_0^{z_{max}} \frac{n(z) F_{\lambda}\Big(\frac{\lambda}{1 + z}\Big)}{(1+z)^3 H(z)}dz
\end{equation}
where $n$ is the comoving density of ALPs, $H(z) = H_0 \sqrt{\Omega_{\Lambda} + \Omega_m (1 + z)^3}$, $H_0$ is the Hubble constant, $c$ is the speed of light and $F_{\lambda}(\lambda)$ is the spectral energy distribution of photons from decaying particles. $\Omega_{\lambda}$ and $\Omega_m$ are the matter and dark energy density parameters (we use the values $\Omega_{\Lambda} = 0.691$, $\Omega_m = 0.309$, $H_0 = 67.8$~km/s/Mpc derived by  \citet{2018arXiv180706209P}). 

If the decay time is much longer than the age of the Universe, or $\Gamma\ll H_0$, $n(z)\simeq const$ and  the integral in Eq. (\ref{eq:i}) could be explicitly taken. This gives the spectral shape of the dark matter decay continuum at a function of $\lambda$:
\begin{equation}
\label{eq:ii}
	\lambda I_{\lambda}(\lambda) = \frac{c}{256\pi^2} \frac{g_{a\gamma\gamma}^2 m^3 \rho_c f \Omega_{dm}}{\Big(\frac{\lambda m}{4\pi}\Big) H_0\sqrt{\Omega_{\lambda} + \Omega_m \Big(\frac{\lambda m}{4\pi}\Big)^3}}
\end{equation}
for $\lambda \ge \lambda_0$ and zero for $\lambda < \lambda_0$. Here $\rho = \rho_c f \Omega_{dm}$ is the dark matter energy density, $\rho_c$ is the critical energy density, $\Omega_{dm}\simeq \Omega_m$ dark matter density parameter and $f$ is the fraction of dark matter in the form of ALPs. 

Fig. \ref{fig:eblspec} shows a compilation of the measurements of the EBL flux using different techniques. The discrepancy between the direct measurements and estimates of the EBL flux based on \gr\ observations is evident particularly in the wavelength range 1-3 $\mu$m. 
Possible  additional ALP delay flux contributions to the EBL, given by Eq. \ref{eq:ii} are shown by the red shading  in Fig. \ref{fig:eblalp} for the ALP mass and two-photon coupling range which is consistent with the existing $\gamma$-ray data (as detailed in the following sections). One could see that this contribution is also consistent with the direct measurements.

\section{Effect of the ALP decay EBL spectral feature on $\gamma$-ray spectra}

The shape of the spectral feature in the EBL produced by ALP decays is different from a generic log-Gaussian bump shape considered by \citet{eblbump}. Nevertheless, it is a feature of the width $\Delta\lambda\sim \lambda$, so that generic conclusions of analysis of \citet{eblbump} apply also for the ALP decay spectral feature. 

In particular, the analysis of \citet{eblbump} has found that the blazar 1ES $1218+304$ provides the best constraints, due to the availability of deep  VERITAS exposure of the source and high source flux in Fermi/LAT energy band. Spectral analysis of this source limits the amplitude of possible bumps in the EBL spectrum to be at most at the level of 10-15 nW/m$^2$sr on top of the EBL flux produced by the star formation process (also at the level of $\sim 10$ nW/m$^2$sr in the micron wavelength range). Considering this fact, we refine the analysis of \citet{eblbump} for the spectrum of 1ES 1218-304 in the specific case of ALP induced EBL spectral feature, by re-fitting the spectrum with account of attenuation of the $\gamma$-ray flux by the spectral feature with the shape described by Eq. (\ref{eq:ii}). 

\begin{figure}
	\centerline{
		\includegraphics[width=1\linewidth]{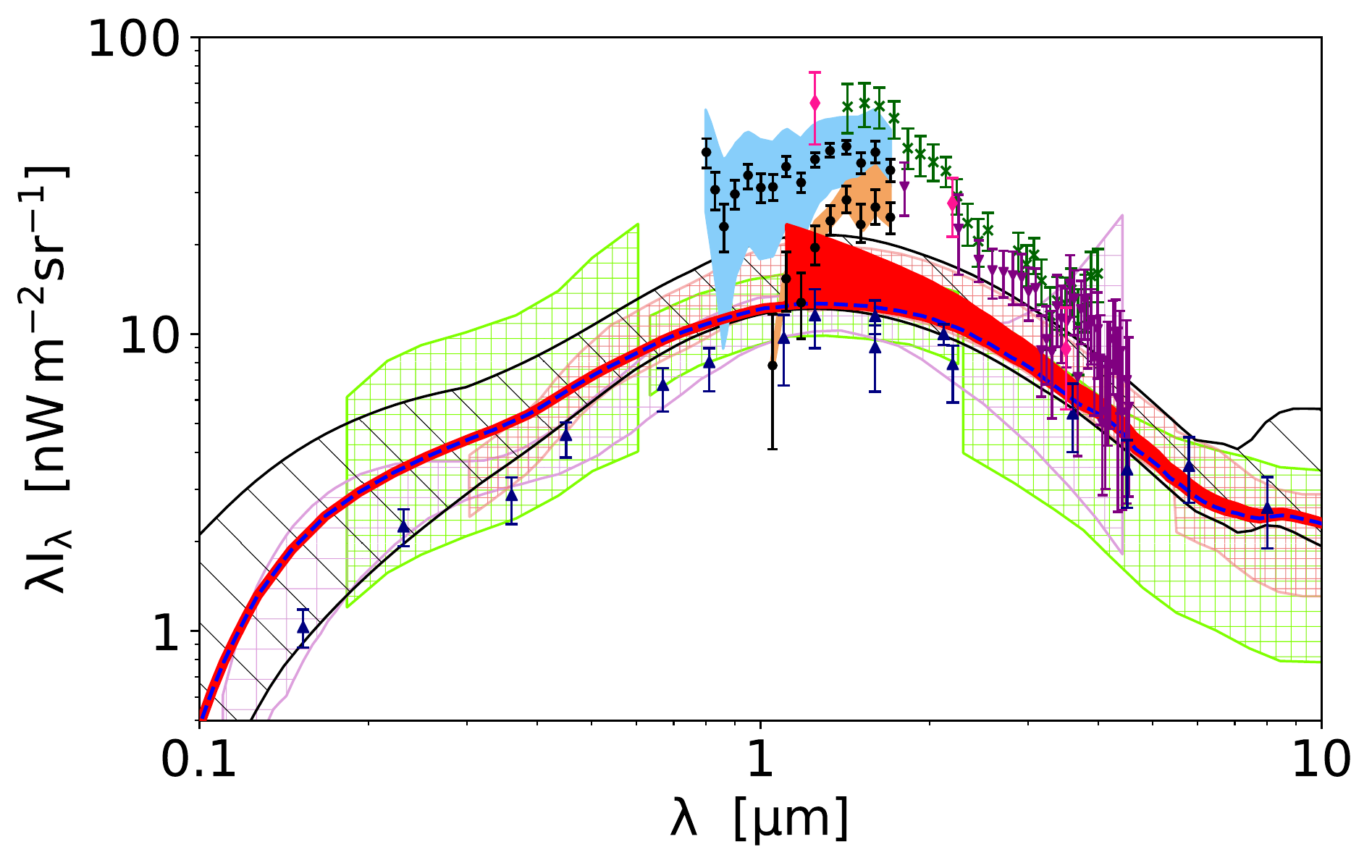}
	}
	\caption{\label{fig:eblspec} Spectral energy distribution of the EBL. 
				 {\it Direct measurements:} purple arrows are results from AKARI
				 \citep{Tsumura:2013iza}, while green asterisks are from the 
				 reanalysis of IRTS \citep{Matsumoto:2015fma}. Pink diamonds are
				 from reanalysis of COBE/DIRBE measurements \citep{Sano:2015bsa,Sano:2015jih}. 
				 Black data points together with blue and orange
				 systematic uncertainty are derived from CIBER \citep{Matsuura:2017lub}
				 and correspond to nominal and minimum EBL models.
				 {\it Lower limits}: dark blue upward arrows combine EBL lower limits
				 obtained by different experiments: GALEX \citep{Xu:2004zg}, Hubble 
				 Deep Field \citep{Madau:1999yh}, Subaru \citep{Keenan:2010na},
				 Spitzer/IRAC \citep{Fazio:2004kx}. {\it EBL from $\gamma$-ray 
				 absorption}: striped lime, red and purple bands are from MAGIC 
				 \citep{Acciari:2019zgl}, HESS \citep{Abramowski:2012ry} and Fermi/LAT
				 \citep{Ajello:2018sxm} correspondingly.
				 {\it Models}: dashed blue line is for baseline EBL model of 
				 \citet{Gilmore:2011ks}. Black striped band shows the allowed range of EBL models 
				 obtained with the global fit to EBL by
				 \citet{Korochkin:2017plb}. Red shaded region
				 shows the favoured range of additional component due to the ALP decay with parameters in the red shaded region  of Fig. \ref{fig:eblalp}.}
\end{figure}

\begin{figure}
	\centerline{
		\includegraphics[width=1\linewidth]{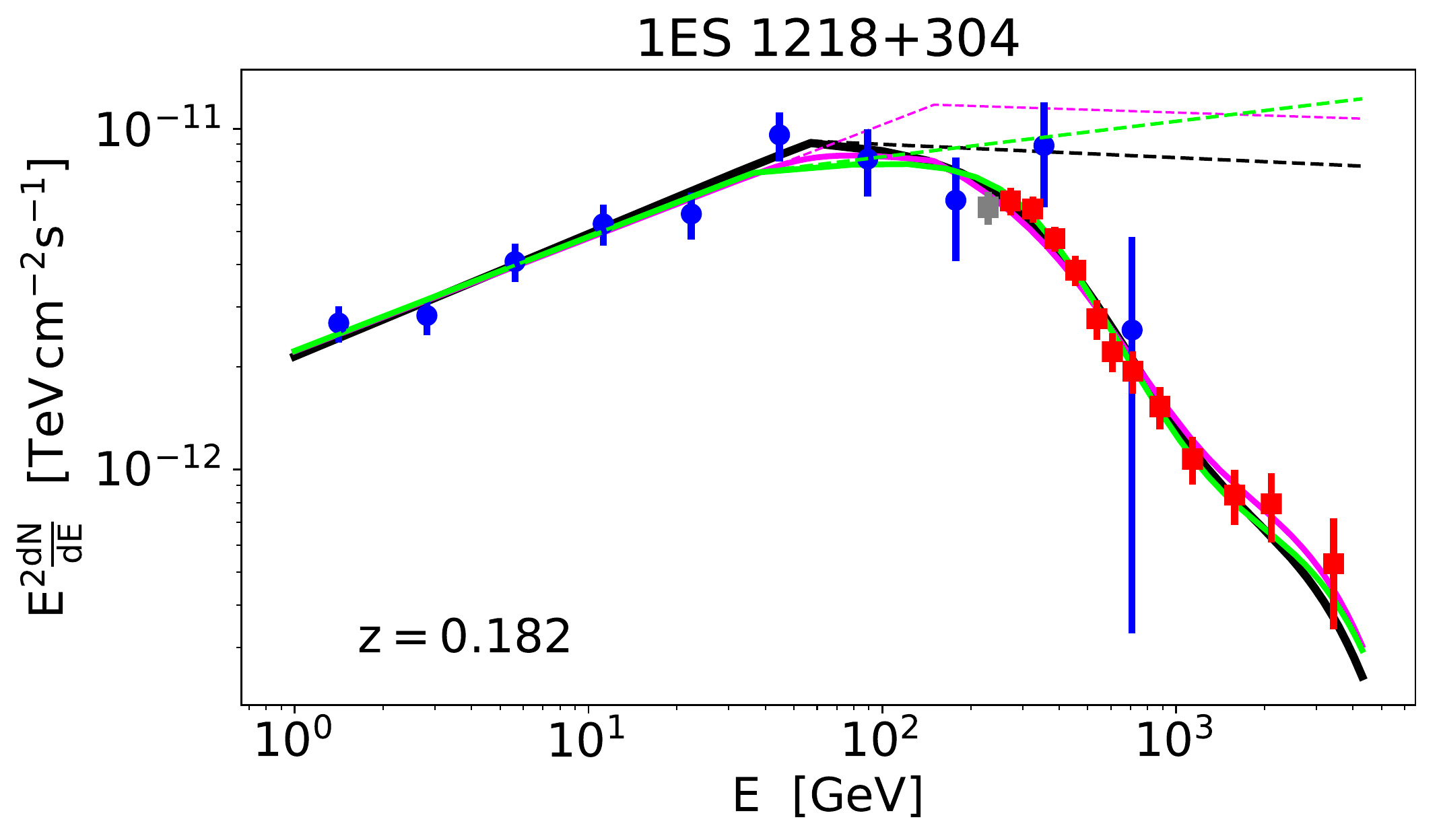}
	}
	\caption{\label{fig:spec} Spectrum of 1ES $1218+304$ measured by Fermi/LAT 
			and VERITAS and theoretical models which fit it with and without axion-like 
			particles. Blue and red points correspond to Fermi/LAT 10 years and 
			VERITAS 86 hours measurements correspondingly. Black dashed thin and black 
			solid thick lines are the best-fit broken power law intrinsic spectra and
			observed spectra absorbed with the baseline EBL model. Dashed and solid 
			green and magenta lines show the intrinsic source and absorbed spectra for 
			the baseline EBL with two additional narrow bumps form axion decays. Grey 
			data point of VERITAS at 250 GeV affect range of allowed parameters 
			presented in Fig. 2 for high axion masses. }
\end{figure}

The best fit model is shown in Fig. \ref{fig:spec}. The intrinsic source spectrum is modelled as a broken powerlaw(s) shown by the dashed lines. The energy of the break found in the spectral fit depends on the assumptions about the EBL model. If the baseline EBL model without additional spectral feature produced by the ALP decay is used, the break is just below 100 GeV energy and the intrinsic spectrum above the break is nearly $dN/dE\propto E^{-2}$. Inclusion of an additional spectral feature in the EBL at $\sim 1 \mu$m shifts the break energy up to 300 GeV and favours softer spectrum above the break (green dashed line). An additional feature in the EBL at $\sim 5 \mu$m shifts the break toward lower energy and favours harder spectrum above the break (magenta dashed line). 

Overall, the spectral fit to 1ES 1218+304 spectrum prefers the EBL model with non-zero bump at approximately one micron wavelength (the effect is at significance level below $3\sigma$ and we could not claim an "evidence" in this sense). The spectral fit corresponding to this best-fit model is shown by the green solid line in Fig. \ref{fig:spec}.

If the constraints on the amplitude of possible bumps in the EBL spectrum are interpreted in terms of the model of decaying ALPs, they could be converted into constraints on the mass-dependent ALP particle coupling to photons using Eq. (\ref{eq:ii}). The result is shown in Fig. \ref{fig:eblalp}. The fact that the fit with non-zero bump feature is favoured by the data has two effects on the ALP coupling constraint. First, it relaxes the overall upper bound on $g_{a\gamma\gamma}$ in the ALP mass range 1-10 eV. This is clear from comparison of the 95\% confidence level upper limit derived from the 1ES 1218+304 analysis (grey region in Fig. \ref{fig:eblalp}) with the order-of-magnitude estimate of constraint on $g_{a\gamma\gamma}$ from absence of features in the EBL derived by \citet{Arias:2012az} (black dashed line).  Next, a preferred range (68\% confidence level) of ALP mass-coupling parameter space (violet shaded ellipses) appears. It spans a wide range of mass between 0.2 and 20 eV and the coupling range $3\times 10^{-11}$~GeV$^{-1}<g_{a\gamma\gamma}<2\times 10^{-10}$~GeV$^{-1}$, just at the border of the stellar evolution constraint shown by the red horizontal line. 

The favoured region of the parameter space has two domains: a wider ellipse around the central value $m\simeq 3$~eV and $g_{a\gamma\gamma}\simeq 10^{-10}$~GeV$^{-1}$ and a more narrow ellipse stretching up to 20~eV in the ALP mass. This structure of the favoured region is largely determined by the presence of features in the spectrum of 1ES 1218+304 visible in Fig. \ref{fig:eblspec}. The noticeable curvature of the spectrum in 0.3-1~TeV energy range favours the fit with an ALP-induced dip in the spectrum  for the ALP mass $m\simeq 3$~eV. Curvature of the VERITAS spectrum close of its low-energy end in the 0.1-0.3 TeV range favours the presence of the dip in this energy range. This dip would be produced by ALPs with masses in 10~eV range. 

It is possible that the VERITAS spectral measurement at the low-energy end is affected by the systematic effects. To illustrate strong dependence of the shape of the favoured range of parameter space on such systematic effects, we have re-done the spectral analysis after removing the first data point of VERITAS (shown in grey colour in Fig. \ref{fig:spec}). The result is shown in Fig. \ref{fig:eblalp} by the red-hatched region. One could see that removal of the spectral measurement possibly affected by the systematic effect changes the shape of the favoured region: the two ellipses merge into a single favoured region. 

\section{Possible improvement of ALP search sensitivity with dedicated observation strategy and with next generation gamma-ray telescopes}

The effect of removal of one spectral measurement point discussed in the previous section shows the limitation of our analysis, which is dominated by the measurement of single source with single Cherenkov telescope. Better constraints on ALP coupling or, ultimately, detection of the ALP induced bump in the EBL spectrum has to be done using observations of multiple sources with multiple telescopes. 

\begin{figure}
	\centerline{
		\includegraphics[width=1\linewidth]{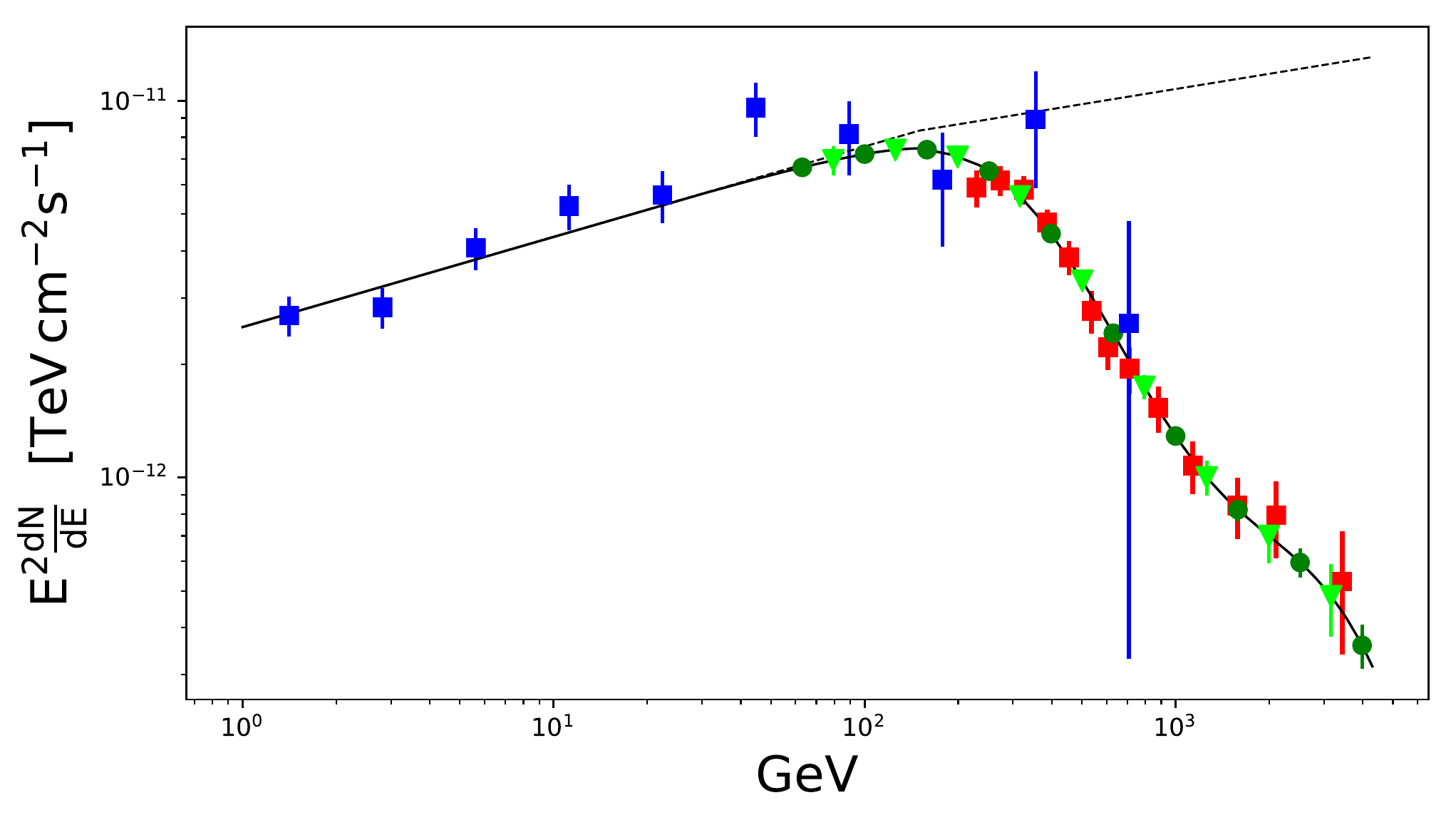}
	}
	\caption{\label{fig:sim_spectra} Simulated spectra of 1ES $1218+304$ for 50~hr long exposures of MAGIC (lime triangles) and CTA North (green circles) telescopes, compared to the VERITAS (red squares) and Fermi/LAT (blue squares) data points. Black think and thick lines show the intrinsic and absorbed spectral models used in the simulations. }
\end{figure}

\begin{figure}
		\includegraphics[width=1\linewidth]{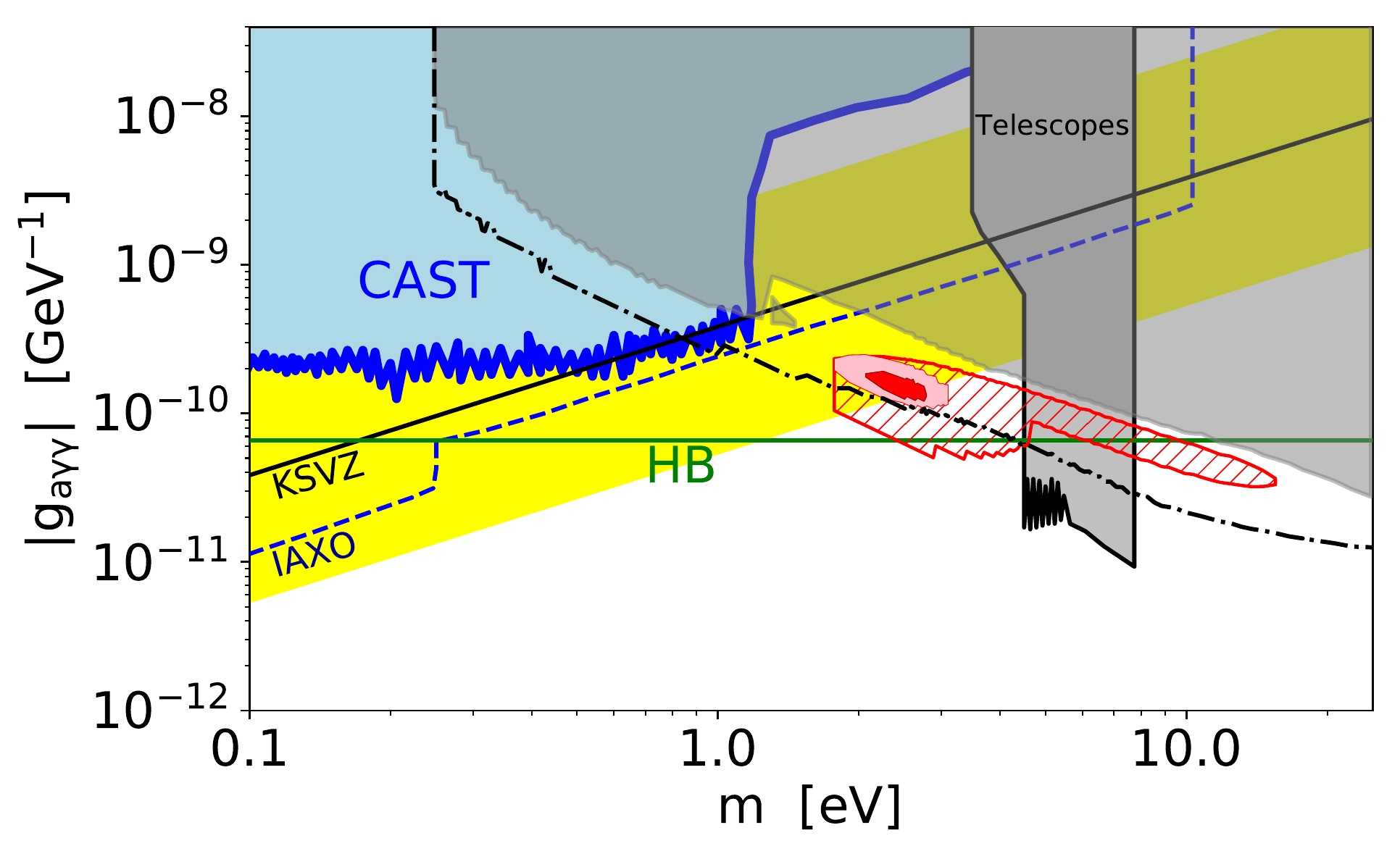}
	\caption{\label{fig:cta} Same as in Fig. \ref{fig:eblalp}, with the sensitivity reach of CTA shown by the black dashed line (95\% confidence level). Pink shaded ellipse shows the preferred ALP parameters (68\% confidence level) obtained with combined VERITAS and simulated MAGIC spectrum of $1218+304$ (50~hr long exposure). Red shaded ellipse shows the precision of the ALP mass and two-photon coupling which can be reached with a 50 hr long exposure of CTA.}  
\end{figure}

This is illustrated in Fig. \ref{fig:cta} where we show the sensitivity reach of observations with MAGIC telescope which have an advantage of lower energy threshold compared to VERITAS. To estimate MAGIC sensitivity for the search of the ALP-induced absorption feature we have simulated the spectral points for a 50 hr MAGIC exposure using the information on the energy dependence of the effective area and background rate found by \citet{magic_crab}. We have then calculated the signal and background statistics and estimated the statistical and systematic errors for each spectral data point based on this information. For the systematic error we have assumed that it is at the level of 5\% of the background flux. The MAGIC spectrum simulated in this way is shown in Fig. \ref{fig:sim_spectra}. 

As it is discussed in the previous section, the spectral fits of Fermi/LAT$+$VERITAS data suffer from uncertainty of the energy of the break in the intrinsic spectrum, which is in the 100 GeV range, but is not well constrained by the data. Lower energy threshold of MAGIC observations would provide a possibility to better constrain the break energy in the 100 GeV range. 

Still further improvement will be achieved with the CTA North array. In order to estimate the sensitivity reach of CTA we have followed the same procedure as for the MAGIC spectral points simulations. We have used the information on the effective area, and background rate within the angular resolution expected for  CTA, available through the CTA website, \url{https://www.cta-observatory.org/science/cta-performance/#1472563453568-c1970a6e-2c0f}, and simulated the expected spectrum of 1ES 1218+304 for a 50 hr observation with CTA North array. The result is presented in Fig. \ref{fig:sim_spectra}. We have chosen the spectral model without an additional ALP induced absorption feature in the spectrum. We have then applied the same analysis as for the VERITAS+Fermi/LAT data fitting the spectrum with spectral model which includes the ALP induced absorption feature, to find the maximal possible normalisation of the  ALP-induced bump in the EBL spectrum as a function of the ALP particle mass. We have then re-calculated the upper limit on the norm of the ALP-induced bump into the upper limit on $g_{a\gamma\gamma}$, to find the sensitivity reach of CTA. The result is shown in Fig. \ref{fig:cta}. From this figure one could see that observations of CTA will provide a sensitivity improvement by a factor of three to five in the ALP mass range around between 0.3 and 10 eV.  

We have repeated the simulation of the 1ES 1218+304 spectrum assuming the EBL model with an additional ALP induced bump at the level of the best fit model of VERITAS+Fermi/LAT spectrum to explore the significance level of detection of the ALP-induced spectral bump in the EBL with the flux consistent with the estimates from the direct EBL measurement experiments \citep{Wright:1999zb,Matsumoto:2004dx,Tsumura:2013iza,Matsumoto:2015fma,Matsuura:2017lub}. We find that if the flux of the ALP induced bump in the EBL spectrum reaches $\sim 10$~nW/(m$^2$s sr), CTA will be able to detect the associated dip in the \gr\ spectrum at significance level $\gtrsim 4\sigma$. Fig. \ref{fig:cta} also shows the precision with which detection of the ALP induced absorption feature with CTA would allow to measure the ALP particle parameters $g_{a\gamma\gamma}$ and $m$. The error ellipse of such measurements is shown by red shading.

\section{Conclusions}

In this work we found 95 \% C.L. limit on axion-like particle mass and coupling constant (see Fig. \ref{fig:eblalp}) from non-observation of axion decay bump in EBL spectrum, using gamma-ray spectrum of  observations of the blazar 1ES 1218+304 by Fermi/LAT and VERITAS telescopes. Our best fit model favours non-zero amplitude of the ALP induced absorption feature in the \gr\ spectrum. This results in the presence of a favoured range of ALP photon coupling and mass shown by the red shaded or hatched regions in Fig. \ref{fig:eblalp}. We have discussed available possibilities to scrutinize this result with complementary observations with current and next generation ground-based \gr\ telescopes, MAGIC and CTA. CTA will provide a factor of three-to-five improvement of sensitivity for the ALP decay feature searches (Fig. \ref{fig:cta}). If the EBL spectrum has an ALP-induced bump with the flux at the level suggested by the direct EBL measurement experiments, CTA will be able to detect the ALP induced bump with high significance, and provide precision measurement of the ALP mass and photon coupling constant with precision which is indicated by the red shared ellipse in Fig. \ref{fig:cta}.

\section*{Acknowledgements}

AK's work on the analysis of gamma-ray data 
and EBL modelling was supported by the
Russian Science Foundation, grant 18-12-00258. 
AK's stay in the APC laboratory was provided by the 
scholarship "Vernadsky" of the French embassy in Russia.

\bibliographystyle{aa}
\bibliography{ebl_axion}

\end{document}